\documentclass[letterpaper, 10 pt, conference]{IEEEtran}  

\IEEEoverridecommandlockouts                              

\usepackage{graphics} 
\usepackage{comment}
\usepackage{subfigure}
\usepackage{multirow}
\usepackage{multicol}
\usepackage{url}

\usepackage{algorithm}
\usepackage{algpseudocode}

\usepackage{cite}
\usepackage{xfrac}

\usepackage{blindtext, graphicx}
\usepackage{textcomp}
\usepackage{booktabs}
\usepackage{subfig}
\usepackage{amsmath}
\usepackage{makecell}

\title{\vspace{-0.5in}
\LARGE 
Non-Invasive Reverse Engineering of Finite State Machines \\Using Power Analysis and Boolean Satisfiability\vspace{-0.2in}
}

\author{\IEEEauthorblockN{Harsh Vamja, Richa Agrawal and Ranga Vemuri}
\IEEEauthorblockA{
Digital Design Environments Laboratory, Department of Electrical Engineering and Computer Science\\
College of Engineering and Applied Science, 
University of Cincinnati, Cincinnati, Ohio USA\\
Email: \{vamjahs, agrawara\}@mail.uc.edu, vemurir@ucmail.uc.edu\\
}\vspace{-0.3in}}

\usepackage[a4paper, total={7.25in, 10in}]{geometry}
\begin{document}

\begin{titlepage}

{\Huge IEEE Copyright Notice}\\

$\copyright$ 2019 IEEE. Personal use of this material is permitted. Permission from IEEE must be obtained for all other uses, in any current or future media, including reprinting/republishing this material for advertising or promotional purposes, creating new collective works, for resale or redistribution to servers or lists, or reuse of any copyrighted component of this work in other works.\\
\\

{\Large Accepted to be Published in: \textbf{Proceedings of the 2019 IEEE International Midwest Symposium on Circuits and Systems,} August 4-7, 2019, Dallas, Texas, USA.}

\end{titlepage}

\textheight=10.3in

\maketitle
\thispagestyle{empty}
\pagestyle{empty}

\begin{abstract}
In this paper, we present a non-invasive reverse engineering attack based on a novel  approach that combines functional and power analysis to recover finite state machines from their synchronous sequential circuit implementations. The proposed technique formulates the machine exploration and state identification problem as a Boolean constraint satisfaction problem and solves it using a SMT (Satisfiability Modulo Theories) solver. It uses power measurements to achieve fast convergence. Experimental results using the LGSynth'91 benchmark suite show that the satisfiability-based approach is several times faster compared to existing techniques and can successfully recover 90\%-100\% of the transitions of a target machine.
\end{abstract}

\begin{IEEEkeywords}
Black-box Analysis, Finite State Machines, Power Analysis, Reverse Engineering, Satisfiability Checking
\end{IEEEkeywords}

\vspace{-0.1in}

\section{Introduction}
\label{sec:introduction}
\vspace{-0.1in}
Reverse engineering of an integrated circuit (IC) aims to reconstruct a behavioral model of the design implemented in the IC. Destructive reverse engineering is an expensive and tedious process which leaves the IC under test unusable \cite{c10}. In recent years, non-destructive reverse engineering to recover  the functionality of a given IC has gained much interest \cite{c15}.
Non-destructive techniques based on reconstructing the device layer models of the IC by using hi-tech x-ray tomography equipment have been proposed \cite{c16, c17, c18}. They require expensive, sophisticated infrastructure and could be extremely time consuming. Certain black-box functional analysis techniques based on characterizing the machine behavior using only input-output observations have also been proposed.  These usually perform brute-force exploration \cite{c19,c22,c24,c25}. They are relatively inexpensive but focus on extremely small machines due to exponential algorithmic complexity. 

Power Analysis attacks are side-channel attacks which use power consumption values to leak information from the devices. These attacks are non-invasive in nature and use relatively inexpensive equipment \cite{c7}. By observing the power consumption trace of a system with respect to a series of input vectors, it is possible to guess the internal operations or the data being processed.

\begin{figure}[h]
      \centering
      \includegraphics[scale=0.30]{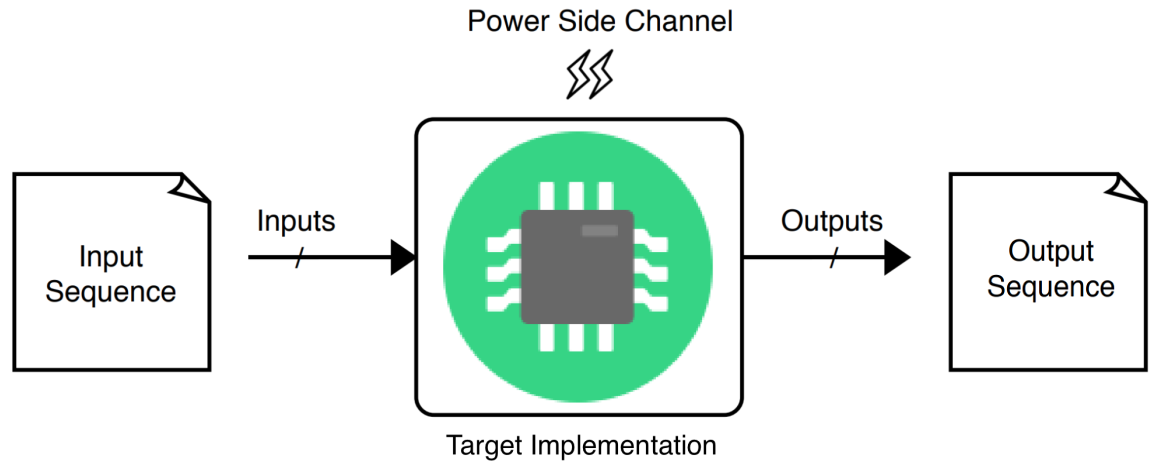}
      \caption{Combined Functional and Power Analysis}
      \label{fig:introduction}
      \vspace*{-0.2in}
\end{figure}

With the explosive growth of IoT devices, smart cards and other small electronic gadgets, it is essential to understand various types of vulnerabilities. In this paper, we propose a non-invasive reverse engineering attack against small-scale digital systems. Using combined functional and power analysis, we propose a method  to recover finite state machines from their synchronous sequential circuit implementations as shown in Figure \ref{fig:introduction}. Combining the two reduces the attack time and memory requirements while increasing the scalability of the attack. 

Section \ref{sec:groundwork} presents the groundwork. We introduce our proposed satisfiability (SAT) solver based FSM recovery method in Section \ref{sec:solver_technique}. Experimental results are presented in Section \ref{sec:experiment_results} and concluding remarks in Section \ref{sec:conclusion}.

\section{Groundwork: HD-Model from Power Analysis}
\label{sec:groundwork}

Let $\mathcal{M}=(I,O,S,\delta,\lambda,s_0)$ be a deterministic finite state machine (FSM) or Moore machine, where $I, O$ and $S$ are finite non-empty sets of inputs, outputs and states respectively, $\delta : I \times S\rightarrow S$ is a state transition function, $\lambda : S \rightarrow O$ is an output function and $s_0 \in S$ is the start-state. 

In sequential circuit implementations of FSMs, states are encoded as Boolean vectors.
Let $B: S \rightarrow (b_1, b_2, \ldots b_R)$ denote a state encoding function where each state is mapped to a Boolean vector of size $R$ and is stored in a state register with $R$ flip-flops.

Let HD(B($s_i$),B($s_j$)) denote the Hamming distance (HD) between two Boolean vectors (B($s_i$) and B($s_j$)) of the same length. 
Circuit implementations in CMOS technology are susceptible to information leakage through power side channels \cite{c7, c30}. The Hamming distance model assumes that dynamic power dissipation in a sequential circuit implemented in CMOS during its transition from state $s_i$ to state $s_j$ is correlated to HD(B($s_i$),B($s_j$)). Given an unknown FSM, we are interested in finding the Hamming distances of the transitions using power analysis attacks in order to discover the state encodings.

First, we perform HD-model based power analysis on {\em known} FSMs to derive a mapping between its transition Hamming distances and the observed power values. This mapping is then used to estimate transition HD values of unknown FSM implementations using power side channel during a reverse engineering attack.

Every FSM state register stores state encoding of the current state of the FSM. During a transition, the contents of the state register get updated which results in power consumption. Hamming distance between these contents should be strongly correlated to its power consumption value. In order to verify the degree of dependency and generate a look-up table to deduce the HD of unknown transitions, sample benchmark machines of varying sizes and connectivity have been tested. 

\begin{figure}[h]
      \centering
          \includegraphics[width=\linewidth]{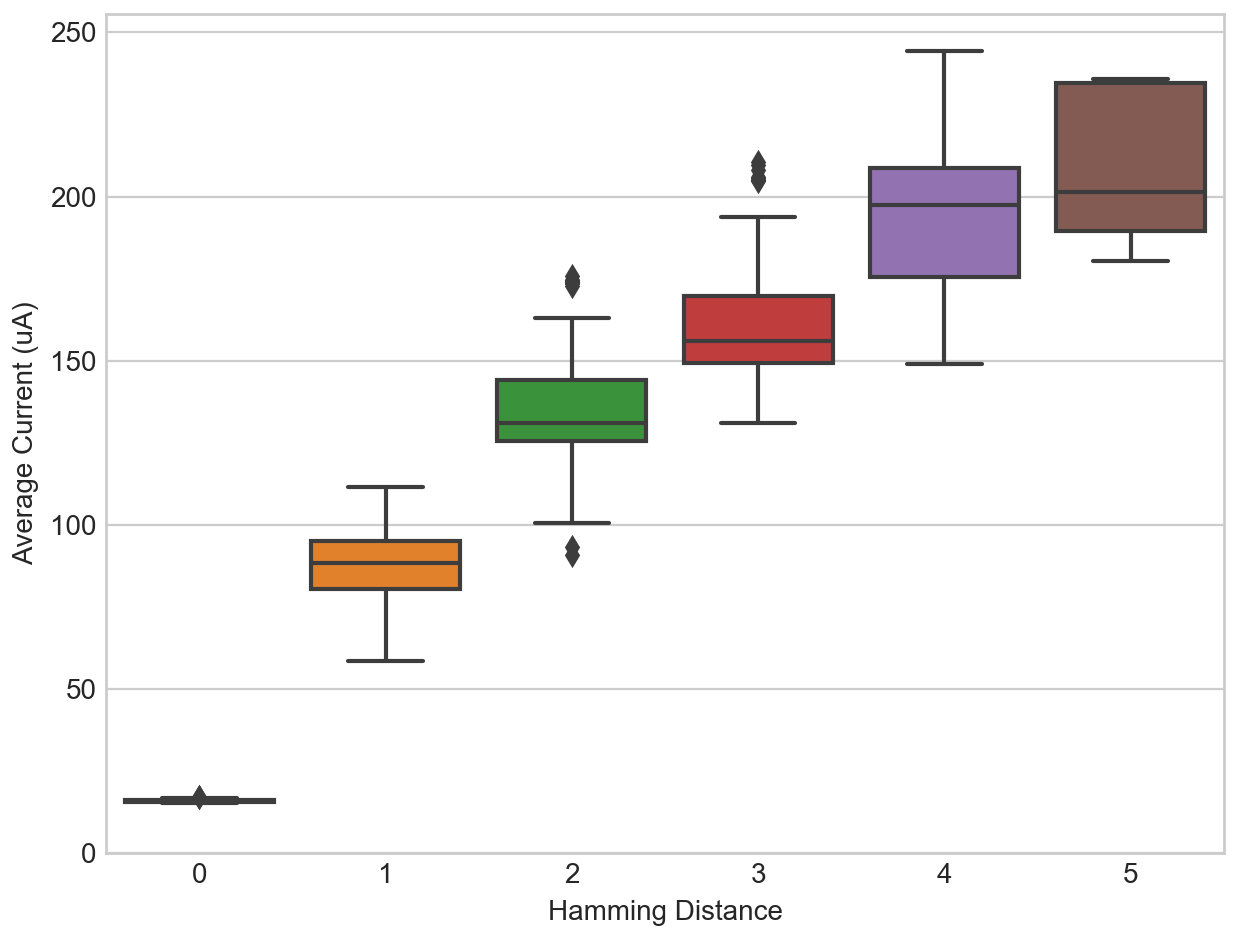}
      \caption{Average Current vs. HD Plot for TBK}
      \label{fig:hd_tbk}\vspace*{-0.2in}
\end{figure}

In order to deduce the relationship between the power values and HD between states, for the SAED90nm CMOS technology, a sample set of LGSynth'91 benchmark FSMs \cite{c35} of varying sizes were tested for varying lengths of input sequences.  Table \ref{tab:Pearson_corelation} shows the Pearson correlation between the HD values and power measurements for 1000 random input vectors. A strong correlation exists between all three statistical measurements of current consumption during transitions and the HD values. In this paper we use average current to infer Hamming distance of transitions. Figure \ref{fig:hd_tbk} shows the average current consumption of 1000 transitions and the corresponding Hamming distance for \textit{TBK} FSM from LGSynth'91 suite. The slight overlap between average current values of consecutive Hamming distances in the figure clearly indicates a possible error of $\pm 1$ during HD inference from power analysis. It is also quite evident that all 0-HD transitions (self-loops) consume the least power and are easily identifiable. These observations are key to justify the $\pm 1$ error in HD-inference and efficiently identify self-loops while trying to reverse engineer the behavior of an unknown machine.

\begin{table}[h]
\centering
{
\begin{tabular}{|l|c|c|c|c|}
\hline
\multicolumn{1}{|c|}{\multirow{2}{*}{{\textbf{Benchmark}}}} & \multirow{2}{*}{\begin{tabular}[c]{@{}c@{}}{\textbf{Transition}}\\ {\textbf{Sample Size}}\end{tabular}} & \multicolumn{3}{c|}{{\textbf{Pearson Correlation Coefficient}}} \\ \cline{3-5} 
\multicolumn{1}{|c|}{} &  & \begin{tabular}[c]{@{}c@{}}{\textbf{Average}}\\ {\textbf{Current}}\end{tabular} & \begin{tabular}[c]{@{}c@{}}{\textbf{Maximum}}\\ {\textbf{Current}}\end{tabular} & \begin{tabular}[c]{@{}c@{}}{\textbf{RMS}}\\ {\textbf{Current}}\end{tabular} \\ \hline
DK15 & 1000 & 0.96795 & 0.92984 & 0.97055 \\ \hline
BEECOUNT & 1000 & 0.94014 & 0.93116 & 0.94269 \\ \hline
BBSSE & 1000 & 0.93645 & 0.89478 & 0.93203 \\ \hline
TBK & 1000 & 0.9444 & 0.95789 & 0.96032 \\ \hline
\end{tabular}
}
\caption{Pearson Correlation Coefficient for Average, Maximum and RMS Current Values}
\label{tab:Pearson_corelation}
\vspace*{-0.2in}
\end{table}

The robustness of the power attack and the reliability of the derived power models can be demonstrated as follows. Power attack is performed on SSE benchmark FSM while treating it as an 'unknown' machine. The attacker can find out it has 7 inputs, 7 outputs and atleast 16 states; and is synthesized using the SAED90nm technology. Testing the unknown machine with 500 randomized input sequences using HSPICE simulation, average current values for all 500 transitions are stored. Table \ref{tab:SPA_Attack_Summary} demonstrates the accuracy of Hamming distances inferred using this power attack. It is observed that 426 transitions out of 500 are inferred correctly and the rest are within the error range $\pm 1$.

\begin{table}[h]{
\begin{center}
\begin{tabular}{|c|c|c|}
\hline
\begin{tabular}[c]{@{}c@{}}{\textbf{Error}}\\ {\textbf{(Inferred HD $-$ Actual HD)}}\end{tabular} & \begin{tabular}[c]{@{}c@{}}{\textbf{No. of}} \\ {\textbf{Transitions}}\end{tabular} & \begin{tabular}[c]{@{}c@{}}{\textbf{Percent of}} \\ {\textbf{Total Transitions}}\end{tabular} \\ \hline
0 & 426 & 85.20\% \\ \hline
1 & 60 & 12.00\% \\ \hline
-1 & 14 & 2.80\% \\ \hline
2 & 0 & 0.00\% \\ \hline
-2 & 0 & 0.00\% \\ \hline
\end{tabular}
\end{center}
\caption{Inferred Accuracy of Power Attack on SSE Benchmark FSM}
\label{tab:SPA_Attack_Summary}\vspace*{-0.2in}
}
\end{table}

By performing similar attacks on a sample of benchmark FSMs \textit{\{dk15, beecount, bbsse, tbk\}} from LGSynth'91 suite, Table \ref{tab:SPA_lookup_table} summarizes the findings in the form of a look-up table. This table can now be used to perform a successful attack on unknown machine and infer Hamming distances of its unknown transitions based solely on its power consumption values.

\begin{table}[h]{
\begin{center}
\begin{tabular}{|c|c|}
\hline
\textbf{Average Current (uA)} & \textbf{Inferred Hamming Distance}\\ \hline
\textless{}40 & 0 \\ \hline
40 to 95 & 1 ($\pm 1$) \\ \hline
95 to 140 & 2 ($\pm 1$)\\ \hline
140 to 170 & 3 ($\pm 1$)\\ \hline
170 to 205 & 4 ($\pm 1$)\\ \hline
205 to 230 & 5 ($\pm 1$)\\ \hline
\textgreater{}230 & 6 ($\pm 1$)\\ \hline
\end{tabular}
\end{center}
\caption{Mapping Between Observed Average Current and Inferred Hamming Distance of State Transitions in 90nm Technology}\label{tab:SPA_lookup_table}\vspace*{-0.2in}
}\end{table}

\section{Boolean Constraint Based Reverse Engineering Attack}
\label{sec:solver_technique}

To perform the attack, random input sequences are used for machine traversal and the output sequences along with the corresponding average power traces are captured. These responses are converted into a set of Boolean constraints, which can be solved using a satisfiability solver.

\subsection{Constraint Formulation for Reverse Engineering}

\subsubsection{Power Analysis Constraints}
Using the lookup table \ref{tab:SPA_lookup_table}, the power trace can be mapped to HD inferences. Table \ref{tab:SPA_Attack_Summary} demonstrated that the inferred HD values are within an error margin of \textit{one}, except for self-loop transitions whose 0-HD values can be precisely identified. Therefore,
\begin{equation}\label{eq:inferredHD}
HD_{actual}-1\leq HD_{inferred}\leq HD_{actual}+1
\end{equation}

\subsubsection{Functional Analysis Constraints}
Output function of the Moore FSM depends on its current state. Therefore, for any two transitions resulting in different outputs, it can be inferred that their resulting states are distinct from one another. On the other hand, identical outputs after transitions, do not necessarily imply identical new states.


\subsubsection{Boolean SAT Formulation}
The problem of generating a logically equivalent state machine can be expressed as a Boolean satisfiability (SAT) problem. Let $N$ input vectors be applied to the target circuit, resulting in $N$ output vectors and $N$ ranges of inferred Hamming distance values as per Equation \ref{eq:inferredHD}:
$$
O = \{o_0, o_1, o_2,...,o_N\}\\
$$\vspace*{-0.2in} $$
HD = \{\{hd_1\pm1\}, \{hd_2\pm1\}, \{hd_3\pm1\},...,\{hd_N\pm1\}\}\\
$$

To discover a binary encoding $B:S\rightarrow\{b_1,b_2,...,b_R\}$ of bit-length $R$, we define a set of constraints on the encoding. We define a predicate \textit{IdenticalStates} for states $s_1$ and $s_2$ being identical (in a self-loop transition) by requiring $HD(s_1, s_2)$ should be equal to zero:
\begin{equation}\label{eq:IdenticalStates}
IdenticalStates(s_1, s_2) := \\ \sum\limits_{r=1}^R(B(s_1)_r  \bigoplus B(s_2)_r) = 0
\end{equation}

Similarly, we define a predicate \textit{InferredHD} based on Equation \ref{eq:inferredHD}, where the Hamming distance of the transition lies within a given range of observed Hamming distances:
\begin{equation}\label{eq:InferredHD}
InferredHD(s_1, s_2) := \\ \sum\limits_{r=1}^R(B(s_1)_r  \bigoplus B(s_2)_r) \in [hd_i \pm 1]
\end{equation}

Non-identical outputs within set $O$ at the end of transitions must imply distinct states. We define predicate \textit{DistinctStates} for states $s_1$ and $s_2$ by requiring the Hamming distance to be a positive integer:
\begin{equation}\label{eq:AreStatesDistinct}
DistinctStates(s_1, s_2) := \sum\limits_{r=1}^R(B(s_1)_r  \bigoplus B(s_2)_r) \geq 1
\end{equation}\vspace*{-0.2in}

For a valid state machine which is logically equivalent to the target machine, we need to find a state assignment with an encoding of length $R$ such that it satisfies all the above constraints. In this research, we have used Z3 SMT Solver \cite{c47} to solve for valid state assignment, since it is a highly efficient solver which has the ability to generate models involving bit-vectors and solve constraints based on them.

\vspace*{-0.1in}
\subsection{Algorithm for Reverse Engineering Attack}
Algorithm \ref{algo:solver-based} shows the process of instantiating and solving the constraints (\ref{eq:IdenticalStates}), (\ref{eq:InferredHD}) and (\ref{eq:AreStatesDistinct}) while progressively increasing $R$. The algorithm finds a valid state encoding for the smallest value of $R$ for which it exits. 

The algorithm initially assumes that every transition results in a new state. Therefore, for $N$ random input vectors, $N$ transitions occur resulting in $N+1$ states. The selection of parameter $N$ is determined based on the number of states and input bits of the target machine (as explained later). As equivalent states are recognized with the help of power analysis and \textit{IdenticalStates} constraint, the states are implicitly merged or \textit{folded}, i.e. the solver provides same encodings to these states. Relations between the other states are also revealed during power analysis which translate to \textit{InferredHD} constraint. Both these constraints are applied in lines (6-12), depending on the inferred Hamming distances. In addition, functional analysis reveals input-output behavior which helps determine distinct states within the unknown machine. Lines (13-17) apply \textit{DistinctStates} constraint after comparing every transition in the observed Output set. Upon finding a satisfiable solution, Lines (18-20) print the solution, else Lines (21-23) increment $R$ by one. $R_{min}$ is determined by the number of unique output values observed during application of the $N$ vectors.

{
\begin{algorithm}

\textit{Inputs}: Inferred HDs \textbf{$HD$} \& Outputs \textbf{$O$} obtained from applying $N$ random vectors\\
\textit{Output}: Logically Equivalent State Machine Encodings
\caption{Generate Logically-equivalent FSM Encodings}
\label{algo:solver-based}
\begin{algorithmic}[1]
\State $R = R_{min}$
\While{$True$}{
			 \State \textbf{initialize} SMT Solver S
             \State \textbf{initialize} State $s_i$, $i \in [0, N ]$: bit-vector of size R
             \State \textbf{set} timeout = 1000000
            \For {\textbf{each} $hd_i$ in $HD$}
            	\If {($hd_i == 0$)}
				\State S $\leftarrow$ Add `IdenticalStates' Constraint
				\Else
				 \State S $\leftarrow$ Add `InferredHD' Constraint
				\EndIf
			\EndFor
            \For {\textbf{each} pair \{$o_i$,$o_j$\} in $O$}
            		\If {($o_i \neq o_j$)}
					\State S $\leftarrow$ Add `DistinctStates' Constraint
					\EndIf
			\EndFor
            
            \If {\textit{(S is Satisfiable)}}
			\State \textbf{print} {State Encoding Solution}  
            \State \textbf{exit}
            \Else
            \State R = R + 1
			\EndIf
}\EndWhile
\end{algorithmic}
\end{algorithm}
}

It should be noted that the encodings generated lead to recovery of a state machine which is  isomorphically equivalent to the implemented one.

\begin{figure}[h]
      \centering
          \includegraphics[scale=0.4]{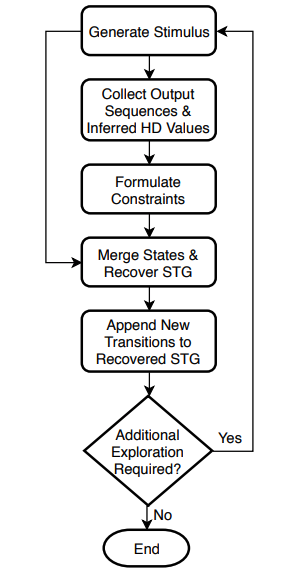}
      \caption{Stimuli Generation Methodology}
      \label{fig:Solver_implementation}\vspace*{-0.1in}
\end{figure}

Selection of an appropriate number of $N$ input vectors is essential to ensure traversal of as many transitions as possible. For a machine with $X$ states and $I$ primary inputs, if the total number of transitions to be recovered is $T$, then $T=X*2^I$. The size of the input vector set is selected to be at least double the value of $T$ so that the algorithm explores that many transitions in one round, hence we choose $N\geq2*T$.  
It is still quite likely that not all transitions would be explored. To cover the missing transitions, the algorithm is repeated with a new set of randomized input vectors to obtain a new state transition graph. By identifying common transitions, based on the input, state and change in output value, the two graphs can be merged to find out new transitions that were not explored in previous rounds. Since every subsequent round will fetch diminishing returns, in our experimental implementation we terminate the process when it recovers 90\% of state transitions from the target machine.  Figure \ref{fig:Solver_implementation} shows the methodology of generating more input vectors as needed. 

\vspace*{-0.1in}
\section{Experimental Results and Analysis}
\label{sec:experiment_results}
\vspace*{-0.1in}
All experiments are performed using FSMs from LGSynth'91 benchmark suite. The machines are translated to Moore machine style by changing the output function while preserving the integrity of all state transitions, state reachability and transition loops. Each FSM is converted to Verilog RTL and synthesized with the Synopsys \textit{SAED90nm} cell library. The resulting gate level netlists are translated to corresponding Spice netlists using a Verilog-to-Spice converter for power and logic simulations. Power traces to perform power analysis are obtained using Synopsys HSPICE and NanoSim. The benchmark machines tested have upto 13 states and 1600 transitions. The number of input bits range from 1 bit to 7 bits and output bits range from 1 bit to 9 bits \cite{c35}.

\begin{figure}[h]
      \centering
        \includegraphics[width=\linewidth]{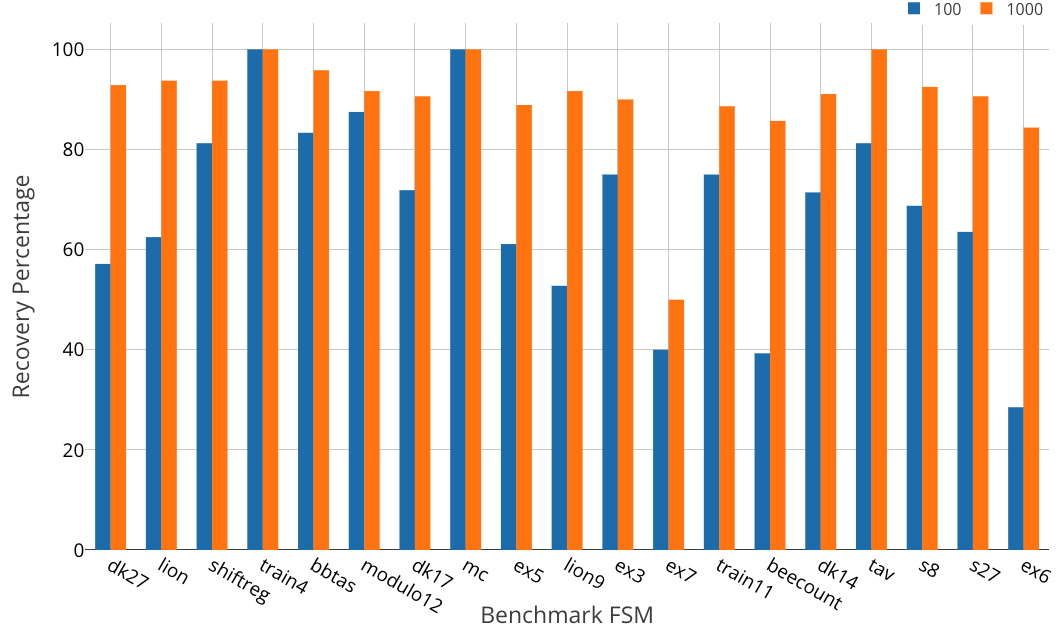}
      \caption{Recovery Percentage in One Round for Test Vector Size 100 and 1000}
      \label{fig:recovery_percentage}
      \vspace*{-0.3in}
\end{figure}

The execution time of the Z3 solver depends on the size of the machine and the number of test vectors. Due to the randomized nature of stimulus selection, a given target machine with the same test vector size will exhibit different runtimes for every round. Hence, we report the average execution time to compare the overall performance on different benchmarks. Table \ref{tab:Solver_Results} summarizes the average run-times for different machines with a set of 100 and 1000 input vectors and Figure \ref{fig:recovery_percentage} shows the recovery percentage at the end of first iteration. The brute-force recovery technique \cite{c24} based on input-output analysis could recover machines with a single input bit and up to 25 transitions in 1 minute, whereas technique \cite{c22} could take several hours and lacks applicability due to the requirement of terminating states. Our technique can handle machines that are 64x larger and also achieve much faster convergence. 

\begin{table}[h]
\centering
{
\begin{tabular}{lcc}
\hline
\multicolumn{1}{|c|}{\multirow{2}{*}{\textbf{FSM}}} & \multicolumn{2}{c|}{\textbf{Average Runtime (s)}} \\ \cline{2-3} 
\multicolumn{1}{|c|}{} & \multicolumn{1}{c|}{\textbf{Test Vectors=100}} & \multicolumn{1}{c|}{\textbf{Test Vectors=1000}} \\ \hline
\multicolumn{1}{|l|}{dk27} & \multicolumn{1}{c|}{0.839} & \multicolumn{1}{c|}{176.275} \\ \hline
\multicolumn{1}{|l|}{lion} & \multicolumn{1}{c|}{0.259} & \multicolumn{1}{c|}{0.884} \\ \hline
\multicolumn{1}{|l|}{shiftreg} & \multicolumn{1}{c|}{0.623} & \multicolumn{1}{c|}{45.652} \\ \hline
\multicolumn{1}{|l|}{train4} & \multicolumn{1}{c|}{0.258} & \multicolumn{1}{c|}{0.833} \\ \hline
\multicolumn{1}{|l|}{bbtas} & \multicolumn{1}{c|}{0.854} & \multicolumn{1}{c|}{146.489} \\ \hline
\multicolumn{1}{|l|}{modulo12} & \multicolumn{1}{c|}{0.71} & \multicolumn{1}{c|}{125.518} \\ \hline
\multicolumn{1}{|l|}{dk17} & \multicolumn{1}{c|}{0.798} & \multicolumn{1}{c|}{132.212} \\ \hline
\multicolumn{1}{|l|}{mc} & \multicolumn{1}{c|}{0.675} & \multicolumn{1}{c|}{39.61} \\ \hline
\multicolumn{1}{|l|}{ex5} & \multicolumn{1}{c|}{0.823} & \multicolumn{1}{c|}{601.05} \\ \hline
\multicolumn{1}{|l|}{lion9} & \multicolumn{1}{c|}{0.653} & \multicolumn{1}{c|}{132.242} \\ \hline
\multicolumn{1}{|l|}{ex3} & \multicolumn{1}{c|}{0.801} & \multicolumn{1}{c|}{102.515} \\ \hline
\multicolumn{1}{|l|}{ex7} & \multicolumn{1}{c|}{0.526} & \multicolumn{1}{c|}{74.304} \\ \hline
\multicolumn{1}{|l|}{train11} & \multicolumn{1}{c|}{0.691} & \multicolumn{1}{c|}{81.086} \\ \hline
\multicolumn{1}{|l|}{beecount} & \multicolumn{1}{c|}{0.586} & \multicolumn{1}{c|}{205.046} \\ \hline
\multicolumn{1}{|l|}{dk14} & \multicolumn{1}{c|}{0.859} & \multicolumn{1}{c|}{101.984} \\ \hline
\multicolumn{1}{|l|}{tav} & \multicolumn{1}{c|}{0.771} & \multicolumn{1}{c|}{70.926} \\ \hline
\multicolumn{1}{|l|}{s8} & \multicolumn{1}{c|}{0.53} & \multicolumn{1}{c|}{45.067} \\ \hline
\multicolumn{1}{|l|}{s27} & \multicolumn{1}{c|}{0.565} & \multicolumn{1}{c|}{40.344} \\ \hline
\multicolumn{1}{|l|}{ex6} & \multicolumn{1}{c|}{1.027} & \multicolumn{1}{c|}{657.988} \\ \hline
\multicolumn{1}{|l|}{bbara} & \multicolumn{1}{c|}{0.86*} & \multicolumn{1}{c|}{179.694} \\ \hline
\multicolumn{1}{|l|}{opus} & \multicolumn{1}{c|}{0.83*} & \multicolumn{1}{c|}{230.357} \\ \hline
\multicolumn{1}{|l|}{ex4} & \multicolumn{1}{c|}{1.2*} & \multicolumn{1}{c|}{301.787} \\ \hline
\multicolumn{1}{|l|}{s386} & \multicolumn{1}{c|}{0.95*} & \multicolumn{1}{c|}{462.514} \\ \hline
\multicolumn{3}{l}{\textit{*Equivalent machine not recovered due to limited exploration}}
\end{tabular}
}
\caption{Average Execution Time for 90-100\% Recovery}
\label{tab:Solver_Results}
\vspace*{-0.2in}
\end{table}

Overall performance of the proposed algorithm depends largely on the number of Z3 solver constraints and how relaxed or tight a given set of constraints are. The following contributing factors are worth noting:

\subsubsection{Self-loops}
The solver will quickly generate a satisfiable model for machines with large number of 0-HD transitions due to its restricted search space. For example, benchmarks \textit{lion}, \textit{train4} \textit{s8} have more than 50\% of their total transitions as self-loops and converge faster than other machines of similar size.

\subsubsection{Number of Primary Outputs and Output Function}
The cardinality of output alphabets for machines having fewer primary outputs will naturally be small and hence, its output function will map multiple states to the same output alphabet. Fewer state pairs with dissimilar outputs lead to a smaller set of state-output based constraints. Due to such relaxed constraints, the solver will converge faster for such machines. This benefit in speed, however, comes at the cost of suboptimal state folding. Benchmarks \textit{train4}, \textit{train11} and \textit{s27} have one primary output and exhibit this behavior, whereas benchmark \textit{ex6} has 8 primary outputs and takes the longest to converge.

\subsubsection{Timeout Parameter}
The proposed algorithm aims to obtain a minimal length state encoding so the solver requires more time to converge with an increase in the number of bit vector variables and constraints. For machines with over 35 states, Z3 fails to generate a minimal length encoding within a reasonable time, but succeeds to produce a model having longer encoding length. Non-minimal state encodings are undesirable as state folding in that case will not be optimal and many indistinguishable states will be misidentified as distinct. 

\vspace*{-0.2in}
\section{Conclusion}
\label{sec:conclusion}
This work proposed a novel approach of combined functional and power analysis to efficiently discover a logically equivalent state machine structure for a target sequential circuit implementation. The proposed technique is faster and scalable to handle larger machines than existing methods. Recovery of 90\%-100\% was achieved for all benchmark FSMs in under 11 minutes. Future work on adaptive input vector generation to perform guided exploration can uncover the remaining transitions for complete recovery.









\end{document}